\documentclass[apjl]{emulateapj}
\usepackage{epsfig}
\usepackage{apjfonts}
\usepackage{amsmath, amssymb,bm}
\begin{document}

\title{Formation of circumbinary planets in a dead zone}

\author{Rebecca G. Martin\altaffilmark{1,4}}
\author{Philip J. Armitage\altaffilmark{1,2}}
\author{Richard D. Alexander\altaffilmark{3}}
\affil{\altaffilmark{1}JILA, University of Colorado \& NIST, UCB 440,
  Boulder, CO 80309, USA}
\affil{\altaffilmark{2}Department of Astrophysical and Planetary Sciences, University of Colorado, Boulder, CO 80309, USA}
\affil{\altaffilmark{3}Department of Physics \& Astronomy, University
  of Leicester, Leicester, LE1 7RH, UK}
\affil{\altaffilmark{4}NASA Sagan Fellow}

\begin{abstract}
Circumbinary planets have been observed at orbital radii where binary
perturbations may have significant effects on the gas disk structure,
on planetesimal velocity dispersion, and on the coupling between
turbulence and planetesimals. Here, we note that the impact of all of
these effects on planet formation is qualitatively altered if the
circumbinary disk structure is layered, with a non-turbulent midplane
layer (dead zone) and strongly turbulent surface layers.  For close
binaries, we find that the dead zone typically extends from a radius
close to the inner disk edge up to a radius of around $10-20\,\rm au$
from the centre of mass of the binary. The peak in the surface density
occurs within the dead zone, far from the inner disk edge, close to
the snow line, and may act as a trap for aerodynamically coupled
solids.  We suggest that circumbinary planet formation may be {\em
  easier} near this preferential location than for disks around single
stars. However, dead zones around wide binaries are less likely and
hence planet formation may be more difficult there.
\end{abstract}


\section{Introduction}

The majority of Sun-like stars form in binary or multiple star systems
\citep{duquennoy91} and circumbinary disks result from their formation
processes \citep[e.g.][]{monin07, kraus12, harris12}. The disks are
similar to circumstellar disks but the central binary provides a
torque on the inner edges of the disk that prevents the majority of
the accretion flow on to the binary. Thus, circumbinary disks resemble
decretion, rather than accretion, disks \citep{pringle81,pringle91}.
These disks are the site for circumbinary planet formation.

There are currently at least thirteen circumbinary planets that have
been observed. About half of these are around evolved stars
\citep[e.g.][]{thorsett93, parsons10,qian10,qian12,qian12b}. There are
six planetary systems with a total of seven planets observed around
main-sequence close binary stars (see Table~1).  The planets in these
systems lie at radii of only a few times the binary separation, where
several physical effects are likely to make planet formation more
difficult than at the equivalent radius in a circumstellar disk.
First, the predicted collision velocities of planetesimals -- if
treated as test particles in a massless disk -- are boosted by the
secular eccentricity imparted from the binary
\citep[e.g.][]{moriwaki04,scholl07,marzari08}.  The equilibrium
eccentricity depends on the magnitude of gas drag, and is strongly
suppressed by precession from even a moderately massive gas disk
\citep{rafikov13}. Nonetheless, estimates suggest that in situ
formation of the observed circumbinary planets is unlikely; the
observed planets must have formed farther out in the disk and migrated
inwards \citep{paard12,rafikov13}. This scenario is supported by
hydrodynamical simulations of the evolution of planetary cores
\citep{pierens07,pierens08}. Second, both the radial and
non-axisymmetric structure \citep{marzari12} of the gas will be
modified by the torque from the binary, altering both the aerodynamic
migration of small solids and the Type~I migration of planetary
cores. Finally, it has been suggested that the interaction between the
binary perturbation and stochastic planetesimal excitation from
turbulence \citep{ida08} may lead to higher equilibrium eccentricities
in circumbinary as compared to circumstellar disks \citep{meschiari}.

If a circumstellar or circumbinary disk is fully ionized, the
magnetorotational instability (MRI) drives turbulence and angular
momentum transport \citep{balbus91}. However, below a critical level
of ionization, a dead zone may form at the midplane where the MRI does
not operate \citep{gammie96,armitage11}. Surface layers of the disk
may be ionized by cosmic rays or X-rays from the central stars
\citep{glassgold04}.  The restricted flow through the disk causes a
build up of material in the dead zone. Moreover, the quiescent dead
zone layer allows solids to settle to the midplane and is thus a
favourable formation site for planets.

In this work, we compute simplified models for poorly ionized
circumbinary disks that contain dead zones. We show that all of the
aforementioned physical processes that impact planet formation in
circumbinary disks are strongly modified in dead zones, and argue that
circumbinary planet formation may even be easier than circumstellar
planet formation at particular disk radii.

\begin{table*}
\centering
\caption{Summary of observations of circumbinary planets around main-sequence stars.}
\begin{tabular}{lccccccccc}
\hline
\hline
Star System& $M_1/\rm M_\odot$ & $M_2/\rm M_\odot$ & $a/\rm au$ & $e_{\rm b}$ & Planet Name & $M_{\rm p}/\rm M_{\rm J}$ & $a_{\rm p}/\rm au$ & $e_{\rm p}$ & Reference\\
\hline
\hline
Kepler-16   & 0.654   & 0.1959  & 0.22   & 0.16 & b &  0.333 & 0.7048 & 0.0069 & 1,2 \\
Kepler-34   & 1.0479  & 1.0208  & 0.22   & 0.52 & b &  0.220 & 1.0896 & 0.18   &3 \\
Kepler-35   & 0.8877  & 0.8094  & 0.18   & 0.14 & b &  0.127 & 0.6035 & 0.042  &3 \\
Kepler-38   & 0.949   & 0.2492  & 0.1469 & 0.103& b &  0.21  & 0.4644 & $<0.032$ &4 \\
Kepler-47   & 1.043   & 0.362   & 0.0836 & 0.0234 & b &$\sim 0.03$  &0.2956& $<0.035$ &5 \\
            &         &         &        &  & c &  $\sim 0.06$    & 0.989  & $<0.411$ &5\\ 
Kepler-64   & 1.528   & 0.408   & 0.18   &  &PH1&  0.531 & 0.56   & 0.1   &6, 7  \\
\hline
\end{tabular}
\tablecomments{Column 2 show the primary star mass, column 3 shows the
  secondary star mass, column 4 and 5 show the semi-major axis and
  eccentricity of the binary orbit respectively, column 7 shows the
  mass of the planet and columns 8 and 9 show the semi-major axis and
  eccentricity of the planet's orbit respectively. } \tablerefs{(1)
  \protect\cite{doyle11}, (2) \protect\cite{bender12}, (3)
  \protect\cite{welsh12}, (4) \protect\cite{orosz12a}, (5)
  \protect\cite{orosz12b}, (6) \protect\cite{schwamb12} and (7)
  \protect\cite{kostov12}.  }
\end{table*}

\section{Circumbinary Disk Model}

In this Section we first describe the model consisting of a binary
star with a coplanar\footnote{If the disk were to form strongly
  misaligned with the binary orbit, the timescale for realignment is
  short, about twenty orbital periods \citep{bate00}. Thus, we would
  only expect significant misalignment in the widest period
  binaries. We only consider models with an aligned binary and disk
  plane.} circular circumbinary disk orbiting the centre of mass of
the stars. Then we describe results from the simulations for various
disk parameters.

\subsection{Model Description}

Previous work on circumbinary accretion disks has assumed the disk to
be fully ionized and turbulent. We follow \cite{alexander12} with the
use of a one-dimensional circumbinary disk model but incorporate a
layered disk model \citep{martin11,martin13a, lubow12,lubow13}. The
disk is thermally ionized, and MRI active throughout, if the midplane
temperature is greater than some critical value, $T_{\rm c}>T_{\rm
  crit}$. The value of the critical temperature is thought to be
around $800\,\rm K$ \citep{umebayashi83}, but we find that this
temperature is not reached within the circumbinary disk. Below this
temperature, the disk may become layered. The surface layers of the
disk (with maximum surface density $\Sigma_{\rm crit}$) are ionized by
cosmic rays or X-rays from the central stars. If the total surface
density is greater than the critical value, $\Sigma>\Sigma_{\rm
  crit}$, then the active layers have surface density $\Sigma_{\rm
  m}=\Sigma_{\rm crit}$ and the midplane contains a dead zone with
mass $\Sigma_{\rm d}=\Sigma-\Sigma_{\rm crit}$. If $\Sigma<\Sigma_{\rm
  crit}$, then the whole disk is externally ionized and MRI active,
$\Sigma_{\rm m}=\Sigma$. This dead zone may become self gravitating if
sufficient mass builds up such that the \cite{toomre64} parameter,
$Q$, is less than a critical value that we take to be 2. However, in
this work, we find that there is not sufficient build up in a
circumbinary disk for self-gravity.

The structure of the MRI active layer and the dead zone has not yet
been well constrained. The surface density that is ionized by cosmic
rays or X-rays, $\Sigma_{\rm crit}$, has been calculated from the
ionization balance of the external effects and internal effects such
as Ohmic and ambipolar diffusion
\citep[e.g.][]{bai11,simon13}. However, these calculations find
accretion rates that are much lower than those observed in T Tauri
stars, that require $\Sigma_{\rm crit}>10\,\rm g\,cm^{-2}$
\citep[e.g.][]{perez11}. With these uncertainties in mind, we allow
$\Sigma_{\rm crit}$ to be a free parameter
\citep[e.g.][]{armitage01,zhu09,zhu10} and consider different values.

Material orbits the central binary of mass $M=M_1+M_2$, at radius $R$,
at Keplerian angular velocity, $\Omega=\sqrt{GM/R^3}$. The governing
accretion disk equation, that includes a binary torque term, for the
evolution of the total surface density $\Sigma=\Sigma_{\rm
  m}+\Sigma_{\rm d}$, and time $t$ is
\begin{align}
\frac{\partial \Sigma}{\partial t} = 
 \frac{1}{R}\frac{\partial}{\partial R} \biggr\{  & 
 3R^{1/2}\frac{\partial}{\partial R}\left[R^{1/2}(\nu_{\rm m}\Sigma_{\rm m}+\nu_{\rm d}\Sigma_{\rm d})\right]  \cr
 &  -\frac{2\Lambda \Sigma R^{3/2}}{(GM)^{1/2}}\biggr\}
\label{main}
\end{align}
\citep{pringle81,lin86}. The viscosity in the MRI active surface
layers is
\begin{equation}
\nu_{\rm m}=\alpha_{\rm m}\frac{c_{\rm m}^2}{\Omega},
\end{equation}
where $\alpha_{\rm m}=0.01$ \citep[e.g.][]{hartmann98} is the
\cite{shakura73} viscosity parameter, $c_{\rm m}=\sqrt{{\cal R} T_{\rm
    m}/\mu}$ is the sound speed, $T_{\rm m}$ is the temperature in the
layer, ${\cal R}$ is the gas constant and $\mu$ is the gas mean
molecular weight. Similarly, the viscosity in the dead zone layer is
\begin{equation}
\nu_{\rm d}=(\alpha_{\rm d}+\alpha_{\rm g})\frac{c_{\rm s}^2}{\Omega},
\end{equation}
where $c_{\rm s}=\sqrt{{\cal R}T_{\rm c}/\mu}$ is the sound speed at
the midplane. The $\alpha_{\rm g}$ term due to self-gravity is zero
unless $Q<Q_{\rm crit}$ \citep[see][for more details]{martin11}. MHD
simulations suggest that there may be some residual viscosity in the
dead zone and this is parameterised with $\alpha_{\rm d}$ but it's
value is still undetermined \citep[e.g.][]{fleming03,simon11}.  We
take $\alpha_{\rm d}=0$ in most of our models but consider one with a
high value of $\alpha_{\rm d}=10^{-4}$ for comparison.

The tidal torque from the binary is
\begin{equation}
\Lambda(R,a)=\frac{q^2 GM}{2R}\left(\frac{a}{\Delta_{\rm p}}\right)^4
\end{equation}
\citep{armitage02}, where $\Delta_{\rm p}={\rm max}(H,|R-a|)$,
$H=c_{\rm s}/\Omega$ is the disk scale height, $a$ is the binary
separation and $q=M_2/M_1$ is the mass ratio of the stars. Coupled
with equation~(\ref{main}) we solve a simplified energy equation
\begin{equation}
\frac{\partial T_{\rm c}}{\partial t}=\frac{2(Q_+-Q_-)}{c_{\rm p}\Sigma}
\end{equation}
\citep{pringle86,cannizzo93}. The disk specific heat is $c_{\rm p}=2.7
      {\cal R}/\mu$.  The local heating is
\begin{equation}
Q_+=Q_{\nu}+Q_{\rm tid}
\end{equation}
where the viscous heating is
\begin{equation}
Q_{\nu}=\frac{9}{8}\Omega^2(\nu_{\rm m}\Sigma_{\rm m}+\nu_{\rm d}\Sigma_{\rm d})
\end{equation}
and the tidal heating term is
\begin{equation}
Q_{\rm tid}=(\Omega_{\rm b} -\Omega) \Lambda \Sigma
\end{equation}
\citep[e.g.][]{lodato09,alexander11}, where the orbital frequency of
the binary is $\Omega_{\rm b}=\sqrt{GM/a^3}$. We also include a simple
prescription for the effects of irradiation from the central stars. We
approximate the irradiation temperature of the disk with
\begin{equation}
T_{\rm irr} =  T_\star\left( \frac{2}{3 \pi}\right)^\frac{1}{4}\left(\frac{R_\star}{R}\right)^\frac{3}{4}
\end{equation}
\citep{chiang97}, where we take $T_\star=4000\,\rm K$ and
$R_\star=2\,\rm R_\odot$. If the midplane temperature of the disk
drops below the irradiation temperature then we assume that the disk
is isothermal and $T_{\rm c}=T_{\rm m}=T_{\rm e}=T_{\rm irr}$
\citep[see also][]{martin12b}.  The cooling rate is
\begin{equation}
Q_-=\sigma T_{\rm e}^4,
\end{equation}
where $T_{\rm e}$ is the surface temperature of the disk and $\sigma$
is the Stefan-Boltzmann constant. The temperature at the midplane,
$T_{\rm c}$, is related to the temperatures in the active layers,
$T_{\rm m}$, and at the surface, $T_{\rm e}$, by considering the
energy balance in the layered model above the midplane and is
described with equations~(6)-(12) in \cite{martin12b}.

The binary torque clears the inner regions of the binary orbit
\citep{artymowicz94,artymowicz96}. It is strong enough to prevent all
accretion flow into the binary region. Hence, the mass of the disk
remains constant in time.  However, some accretion may occur on to the
binary through tidal streams. The maximum average of this is thought
to be about ten per cent of the steady accretion flow
\citep{macfadyen08}. Hence, we also include a model with accretion
on to the binary stars. Material is removed from the inner regions of
the disk at a fraction $\epsilon$ of the steady accretion rate $3\pi
(\nu_{\rm m} \Sigma_{\rm m}+\nu_{\rm g}\Sigma)$.

The initial surface density is chosen to be
\begin{equation}
\Sigma=\Sigma_0 \left(\frac{R_0}{R}\right) \exp\left(-\frac{R}{R_0}\right), ~~R>10a,
\end{equation}
where we take $R_0=15\,a$ and scale the constant $\Sigma_0$ such that
the total mass of the disk, $M_{\rm disk}$, is equal to some constant
in the range $0.01 - 0.1\,\rm M_\odot$ \citep[see
  also][]{alexander12}. This initial condition has little effect on
the structure of the disk at later times. The total mass of the disk,
not the initial distribution of mass, is the important feature.

We take a grid of 200 points distributed evenly in $\log R$ from
$R_{\rm in}=2a$ up to $R_{\rm out}=1000\,\rm AU$. At the inner edge we
have a zero surface density inner boundary condition, but because the
binary torque is strong, there is no material here. At the outer edge
we take a zero radial velocity boundary condition to prevent mass
loss, but we have chosen a radius large enough that no significant
amount of material spreads this far in a reasonable timescale. In the
following section we describe the results of our simulations.

\subsection{Model Results}

\begin{figure*}
\begin{center}
\includegraphics[width=7cm]{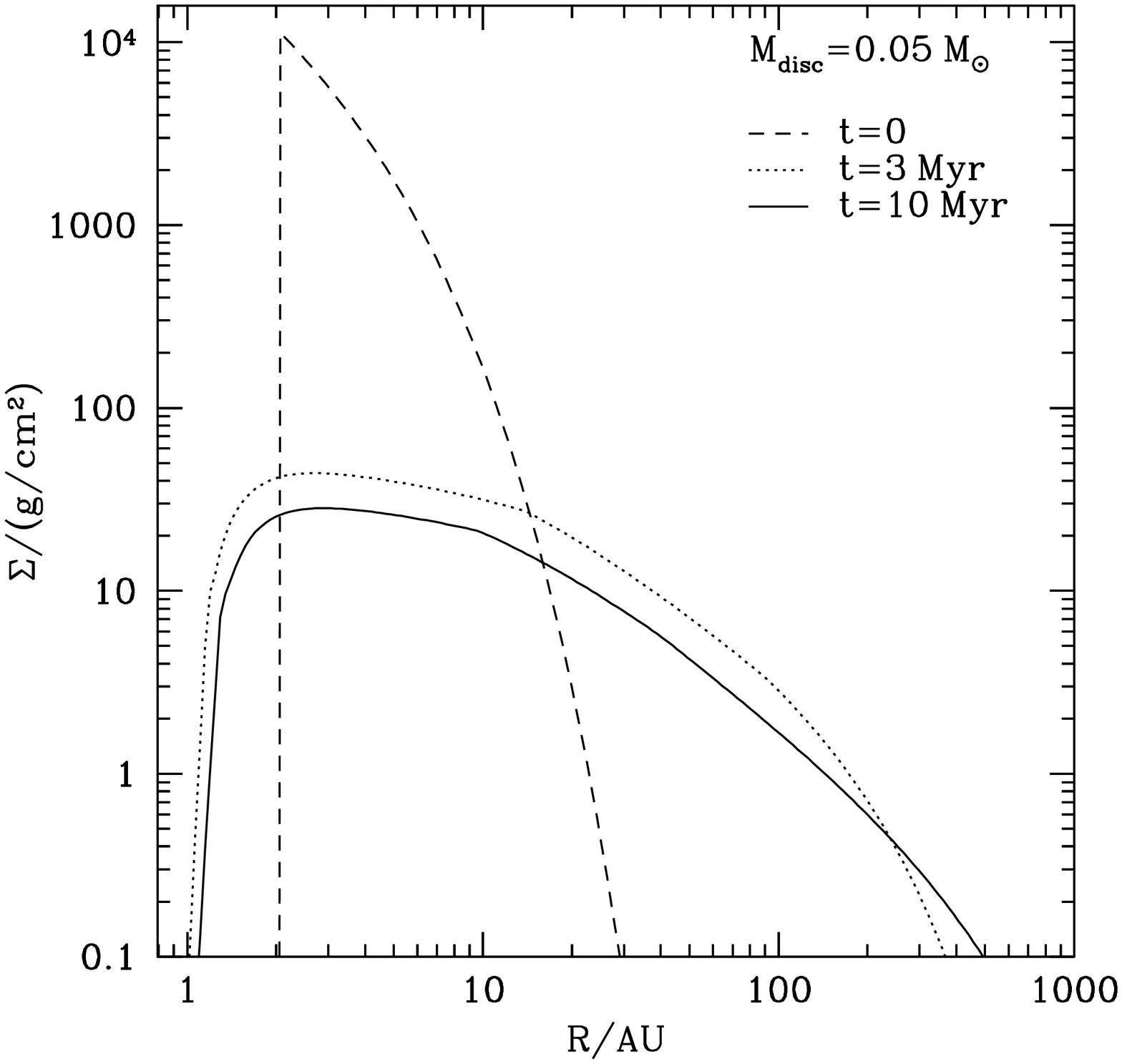}
\includegraphics[width=7cm]{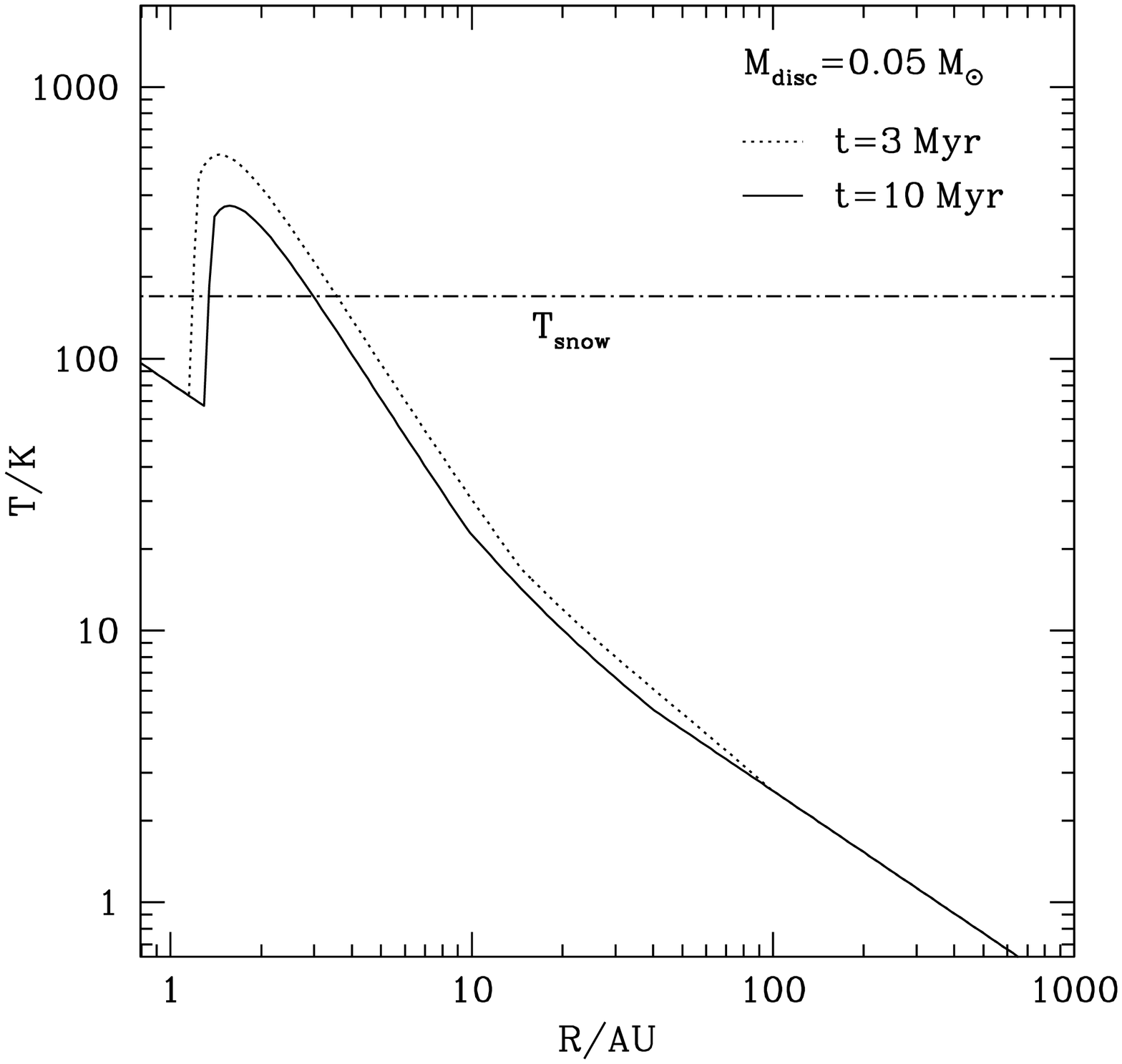}
\includegraphics[width=7cm]{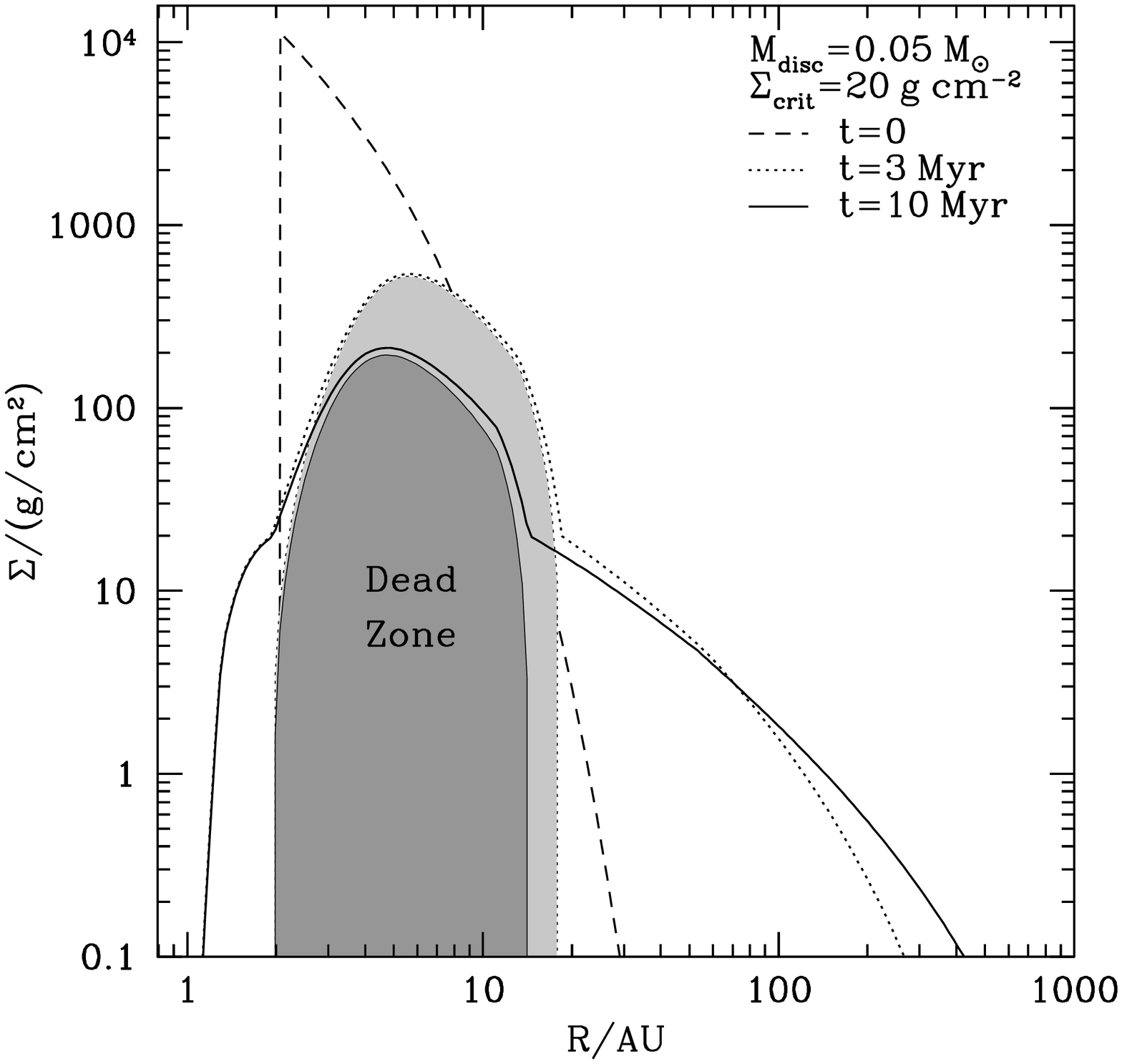}
\includegraphics[width=7cm]{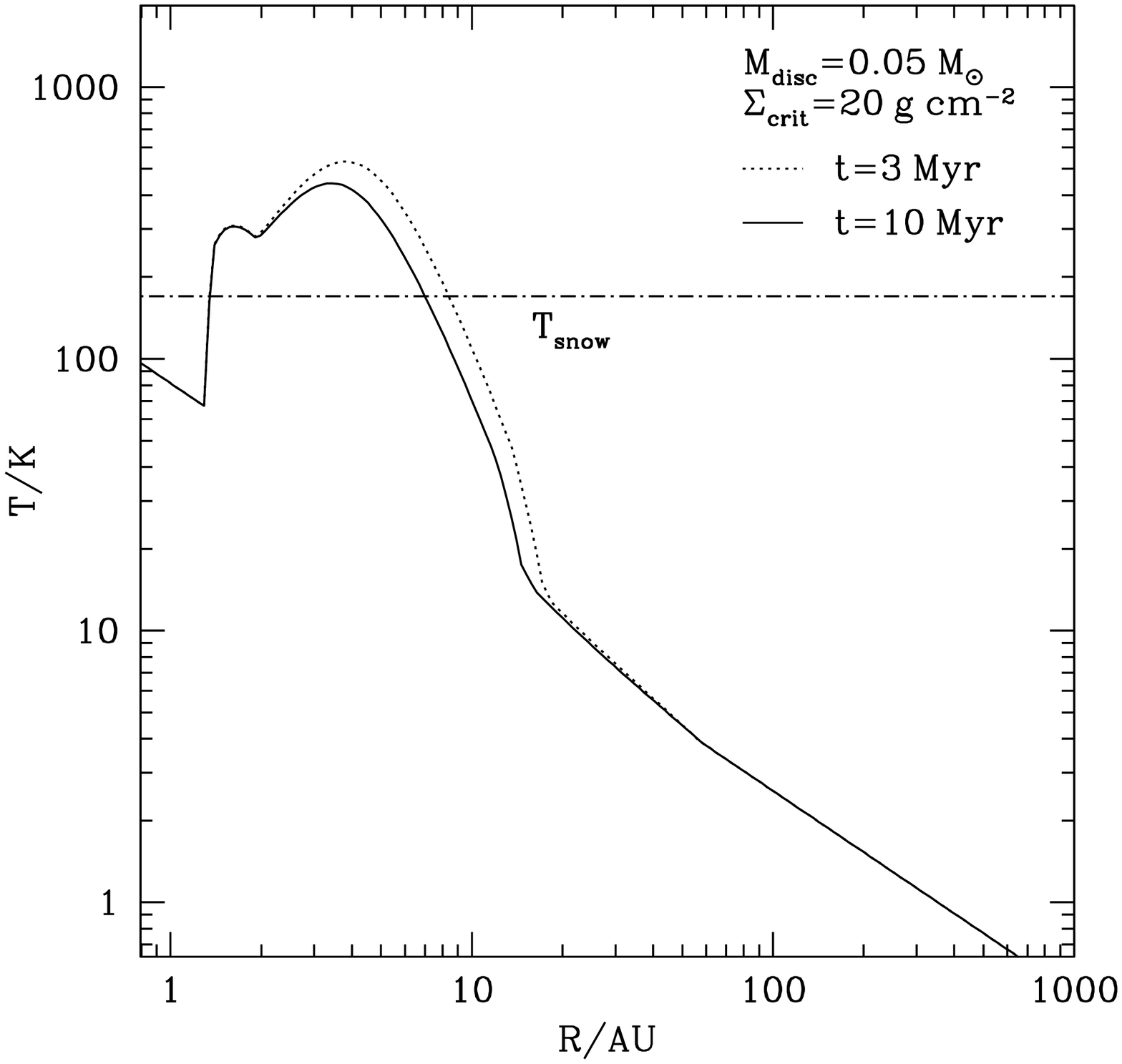}
\end{center}
\caption{The disk surface density (left) and central temperature
  (right) at a time of $t=3\,\rm Myr$ (dotted lines) and $10\,\rm Myr$
  (solid lines) for a fully turbulent circumbinary disk (model R1,
  upper plots) and a disk with a dead zone defined by $\Sigma_{\rm
    crit}=20\,\rm g\,cm^{-2}$ (model R2, lower plots). The dashed
  lines in the surface density plots show the initial surface density.
  The shaded regions shows the surface density of the dead zone at
  $t=3\,\rm Myr$ (pale shaded region) and $t=10\,\rm Myr$ (dark shaded
  region). The dot-dashed lines in the temperature plots show the snow
  line temperature. }
\label{data}
\end{figure*}

Fig.~\ref{data} shows the disk structure for two circumbinary disk
models with central binary star masses of $M_1=1\,\rm M_\odot$ and
$M_2=1\,\rm M_\odot$ with a separation $a=0.2\,\rm au$, accretion
efficiency $\epsilon=0$ and total disk mass $M_{\rm disk}=0.05\,\rm
M_\odot$.  Model R1 is fully turbulent and model R2 has a dead zone
defined by $\Sigma_{\rm crit}=20\,\rm g\,cm^{-2}$. In both cases, the
inner parts of binary region are cleared out to a radius of greater
than $1\,\rm au$.  In model R2, the dead zone structure is set up in
less than $10^5\,\rm yr$ and then is fairly constant in time.  The
disk is fully MRI turbulent in the inner and outer parts where the
surface density is less than the critical that is ionized by external
sources, $\Sigma<\Sigma_{\rm crit}$.  The dead zone extends from
$1.91\,\rm au$ to $18.5\,\rm au$ at a time of $3\,\rm Myr$, for
example, and is not self gravitating at any time. The mass of the dead
zone decreases in time as the disk spreads out in radius. The peak in
the surface density distribution is much farther from the binary star
in the disk with a dead zone, than in the fully turbulent case where
the peak is close to the inner edge of the disk.  The radius of the
peak surface density moves inwards in time. We discuss this feature
further in Section~\ref{peak}.

Similarly, the peak in the temperature distribution of the fully
turbulent disk is closer to the binary star than the disk with a dead
zone.  The snow line radius, $R_{\rm snow}$, is the radius beyond
which ice condenses from the disk, that occurs at a temperature of
around $T_{\rm snow}=170\,\rm K$ \citep[e.g.][]{lecar06,
  martin12,martin13}.  Thus, the snow line temperature occurs at a
radius farther from the binary in the disk with a dead zone. Because
the solid mass density is much larger beyond the snow line,
planetesimal formation is much easier there and thus giant planets may
form more easily.  Hence we expect both gas giants and rocky
terrestrial planets to form in circumbinary disks, as they do in
circumstellar disks. In the dead zone model, terrestrial planet
formation is easier compared with the turbulent model as there is a
larger radius out to which they can form.

In Table~2 we show models with various disk parameters and find that
the dead zone is extensive for a range of disk masses and critical
surface densities. The smaller the critical surface density,
$\Sigma_{\rm crit}$, the larger the dead zone and the larger the
radius of the peak surface density (comparing models R2 and R4, for
example). If the disk mass is small (less than $0.05\,\rm M_\odot$),
the critical surface density must be small, ($\Sigma_{\rm
  crit}<20\,\rm g\,cm^{-2}$) for a dead zone to exist for the whole
disk lifetime, up to $10\,\rm Myr$ (see models R6 and R7). On the
other hand, if the disk is massive, the dead zone is more extensive
and the peak surface density is at a larger radius (see model R8). The
accretion flow from the disk on to the binary (model R5 with
$\epsilon>0$) has little effect on the structure of the disk and the
dead zone.  In this case, the average accretion rate on to the binary
stars is $6.5\times 10^{-10}\,\rm M_\odot\, yr^{-1}$.

\subsection{Peak Surface Density}
\label{peak}

For a disk without a dead zone, the maximum surface density occurs
close to the inner edge of the disk, as shown in the top plots of
Fig.~1. However, in a disk that contains a dead zone, the peak surface
density may lie far from the inner edge (see Table~2).

The inner parts of the circumbinary disk show little radial movement
over the lifetime of the disk (as shown in Fig.~1). In such a disk the
inward or outward movement of dust particles is determined by the sign
of the radial pressure gradient
\citep[e.g.][]{takeuchi02,takeuchi05}. The pressure gradient,
$\partial P/\partial R$, where $P=c_{\rm s}^2\Sigma/\sqrt{2\pi}H$, is
positive inside of the radius of the peak surface density, $R_{\rm
  peak}$, and negative outside. This suggests that particles of all
sizes drift towards and collect around this radius. The radius of the
peak surface density occurs far from the inner edge of the disk for
models with a dead zone. It moves outwards with decreasing active
layer surface density.

\cite{paard12} found that planetestimal accretion can occur in
$R>20a=4.4\,\rm au$ for Kepler-16, $R>12a=2.64\,\rm au$ for Kepler-34
and $R>15a=2.7\,\rm au$ for Kepler-35. The results of \cite{rafikov13}
suggest that these numbers may overstate the difficulty of forming
planetesimals, at least while the gas disk remains relatively
massive. Models R9 and R10 represent the specific example of
Kepler-16. Models R1--R8 represent circumbinary disk models for
Kepler-34 and Kepler-35. For reasonable dead zone models, we find that
the peak surface density is farther out than the region where
planetesimal accretion is inhibited. Thus, the dead zone is a likely
formation site for the circumbinary planets. The binary perturbations
do not affect planetestimal growth in a circumbinary disk with a dead
zone.

\begin{table*}
\begin{center}
\caption{Summary of circumbinary disks models.}
\begin{tabular}{l|cccccc|c|cccc|cccc}
\hline
\hline

Model  & $M_1$  &   $M_2$  & $a$   & $\Sigma_{\rm crit}$ & $\alpha_{\rm d}$  & $\epsilon$ &  $M_{\rm disk}$  & $M_{\rm DZ}$   &  $R_{\rm d1}$ & $R_{\rm d2}$ & $R_{\rm peak}$ & $R_{\rm peak}$ & $M_{\rm disk}$ & $M_{\rm DZ}$ &  $R_{\rm snow}$\\
    &  ($\rm M_\odot$) & ($\rm M_\odot$)  & ($\rm au$) & ($\rm g\,cm^{-2}$)& & & ($\rm M_\odot$) & ($\rm M_\odot$) & (au) & (au) &(au) & (au) & $(M_\odot$)& $(M_\odot$) & (au)  \\
\hline
Time/Myr   & {\rm all} & & & & & & 0 & 3 & & & & 10\\
\hline
\hline
R1 &  1.0 & 1.0  &  0.2   &  $>44$    &0.0 & 0.0  & 0.05   & -  &- &- &2.72 &2.94 &0.05& - & 2.94   \\
R2 &  1.0 & 1.0  &  0.2   &  20    & 0.0 &0.0 & 0.05   & 0.0231    & 1.91 & 18.5  & 5.72  &4.70  & 0.05 & 0.00575 & 5.95 \\
R3 & 1.0 & 1.0 & 0.2 & 20    & $10^{-4}$ & 0.0 & 0.05 & 0.0179 & 1.77 & 20.8 & 4.70 &4.35 & 0.05 & 0.00349 & 5.29 \\
R4 &  1.0 & 1.0 & 0.2 & 10  & 0.0 & 0.0 & 0.05 &0.0369 & 1.91 &21.6 & 7.82 & 7.52 & 0.05 & 0.0249 & 9.15 \\
R5 &  1.0 & 1.0  &  0.2   &  20    & 0.0 & 0.1 &  0.05   & 0.0202  & 1.91 & 18.5 & 5.50 & 4.35 & 0.0435 & 0.00243 & 5.09\\
R6 &  1.0 & 1.0  &  0.2   &  20    & 0.0 & 0.0 &  0.02 & 0.00208 & 1.99 & 11.6 &4.35 & 3.18  & 0.02 & -  & - \\
R7 &  1.0 & 1.0  &  0.2   &  10    & 0.0 & 0.0 &  0.02   & 0.00994 & 1.91 & 17.1 & 6.69 & 5.72 & 0.02 & 0.00328 & 5.72 \\
R8 & 1.0 & 1.0 & 0.2 & 20 & 0.0 & 0.0 &0.1 &0.0669 & 1.70 & 22.5 &  6.43 & 5.94 & 0.1 & 0.0381 & 9.15 \\
\hline
R9   &  0.7 & 0.2 & 0.2  & 20  &  0.0  & 0.0 & 0.05 & 0.0362 & 1.06 & 12.0 & 4.02 & 3.72 & 0.05 & 0.0249 & 5.72\\
R10             & 0.7 & 0.2 & 0.2 & 10 & 0.0 & 0.0 & 0.05 & 0.0435 & 1.20 & 12.0 & 5.29 & 5.09 & 0.05 & 0.0375 & 6.43\\
\hline
R11 &  1.0 & 1.0 & 1.0  & $>13.3$ & 0.0& 0.0  & 0.05   &    & - & - & 12.5 & 12.9 & 0.05 & - & -  \\
R12&  1.0 & 1.0 & 2.0  & $>5.8$ & 0.0 & 0.0 & 0.05  &   - & - & -& 24.7 & 25.4 & 0.05 & -       &  - \\
R13 &     1.0 & 1.0 & 5.0  & $>2.0$ & 0.0 &0.0    & 0.05 & - & - & - & 60.3 &  63.1 & 0.05 & -  &  -\\
\hline
\end{tabular}
\end{center}
\tablecomments{Column 2 shows the mass of the primary star, column 3
  shows the mass of the secondary star and column 4 shows the
  semi-major axis of the binary orbit. Column 5 shows the critical
  surface density that is ionized by cosmic rays or X-rays from the
  central binary and column~6 shows the viscosity parameter in the
  dead zone. Column 7 shows the parameter $\epsilon$ that describes
  the efficiency of accretion on to the binary and columns 8 shows the
  initial mass of the disk. Columns~9-12 describe the disk structure
  at a time $t=3\,\rm Myr$. Columns~9 shows the mass of the dead zone,
  columns~10 and~11 show the inner and outer radius of the dead zone,
  respectively and column~12 shows the radius of the peak surface
  density.  Columns~13-16 describe the disk structure at $t=10\,\rm
  Myr$.  Column~13 shows the radius of the peak surface density,
  column~14 shows the disk mass, column~15 shows the dead zone mass
  and column~16 shows the radius of the snow line. In rows where there
  is no snow line radius shown, the snow line is inside the inner edge
  of the disk. }
\label{table2}
\end{table*}

\subsection{Viscosity in the Dead Zone}

A non-zero viscosity in the the dead zone is uncertain and thus we
have assumed that the dead zone has zero turbulence, $\alpha_{\rm
  d}=0$, in most of our models. However, shearing box simulations
suggest that the MHD turbulence generated in the disk surface layers
may produce some hydrodynamic turbulence in the dead zone layer that
may produce a small but non-zero viscosity
\citep[e.g.][]{fleming03,simon11, gressel12}.

With this uncertainty in mind we also performed one simulation with a
viscosity in the dead zone, model R3. We choose the parameters of
model R2 but include a viscous term in the dead zone with $\alpha_{\rm
  d}=10^{-4}$. We note that this is likely to be an upper limit to the
viscosity in the dead zone.  In a circumstellar accretion disk, the
amount of mass flow through the dead zone should be less than that
through the active layer because the turbulence in the dead zone is
generated by the turbulence in the active layers \citep[see
  also][]{bae13}. Even in this extreme case we find that the disk
structure is not significantly altered. The extent of the dead zone is
actually larger, but the mass in the dead zone is slightly
smaller. The peak in the surface density occurs at a radius slightly
closer to the binary stars, but still farther out than the region that
is hostile to planetesimal formation for Kepler-34 and Kepler-35. If
the magnitude of the viscosity in the dead zone is reduced by a factor
of around ten or more then the effect on our results is minimal. Thus,
we conclude that a small but non-zero viscosity in the dead zone will
not significantly affect the results presented here.

\subsection{Wide Separation Binaries}

The table also includes some circumbinary disk models for larger
separation binaries (models R11 --R13).  Here, the surface density of
the disk (and thus the mass) must be much larger for a dead zone to be
present.  For an equal mass binary with a separation of $1\,\rm au$,
the peak surface density is only $13.3\,\rm g\,cm^{-2}$ (at a time of
$3\,\rm Myr$, see model R11) and this decreases with binary separation
(see for example models R12 and R13). If a dead zone is required for
planet formation, this suggests that circumbinary planet formation is
far more likely in close binaries.

\section{Discussion}

When a massive planet forms within a disk, it can open a gap in the
disk. The gap opening mass, according to the viscous criterion
\citep{lin86}, will be significantly reduced in a dead zone, but the
thermal criterion will be only modestly affected. Even in a normal
circumstellar accretion disk, this mass is well below a Jupiter mass
\citep{zhu13} and for the most massive planets to form, there must be
accretion on to the planet across the gap \citep{lubow06}.  The planet
captures most of the material that flows across the gap and the dead
zone does not drastically affect planetary accretion in a
circumstellar disk \citep{uribe13}.  However, in a circumbinary disk,
there is very little mass flow and thus it is not clear how much
accretion would occur on to a planet in this case. If the accretion is
reduced significantly compared with a circumstellar accretion disk,
then we would expect only lower mass gas giants to form within
circumbinary disks. However, this should be investigated in future
work. Similarly, the presence of the dead zone with a highly reduced
viscosity will mean that migration timescales are significantly
different than in the turbulent disk.

If planets form within the dead zone, thus removing the mass from the
disk, the remaining disk may have a sufficiently low surface density
that the whole disk becomes fully turbulent. Thus, the remaining
turbulent disk may accrete on to the planetary cores allowing the
formation of giant planets in the late stages of disk evolution.

There are a number of improvements that should be made to this disk
model in future, such as modelling the non-axisymmetric tidal streams,
non-axisymmetric stellar irradiation and increasing the eccentricity
or inclination of the binary orbit. The models presented here cannot
model such effects, but are a first step towards understanding the
formation of dead zones in circumbinary disks. If a dead zone is
present in a circumbinary disk, then it remains for the lifetime of
the disk.  This is because the binary torque is preventing the
majority of accretion and the mass of the disk is fairly constant. The
dead zone evolves on a long timescale. However, in the circumstellar
disk, once the infall accretion rate drops, the dead zone may be
accreted and the cosmic rays or X-rays can penetrate the whole of the
less massive disk. This suggests that planet formation in close binary
stars is even more likely in circumbinary disks despite the concerns
about planetesimal accretion in the innermost regions.

A more realistic way to measure the extent of the dead zone may be
with a critical magnetic Reynolds number
\citep[e.g.][]{fromang02,matsumura03}. The active layer surface
density then varies with radius \citep[e.g.][]{martin12a}. However,
models with this prescription cannot account for the accretion rates
observed in T Tauri stars and this issue is yet to be resolved.

We note also that our model does not account for mass-loss due to
photoevaporation.  Photoevaporative winds are thought to drive final
disk clearing in disks around single stars
\citep[e.g.][]{alexander06,gorti09,owen10}.  However,
\cite{alexander12} showed that, because accretion is suppressed by the
tidal torque from the binary, photoevaporation plays a much larger
role in the evolution of circumbinary disks. Photoevaporation steadily
erodes material from near the disk inner edge, and can modify the disk
structure significantly if the mass-loss rate is high
enough. Moreover, for the close binary separations considered here
much of the photoevaporative mass-loss comes from $\sim$AU radii,
similar to the locations of our predicted dead zones. Crudely, we
expect photoevaporative mass-loss to shorten the lifetimes of the dead
zone, and perhaps also to modify the radial surface density profile at
radii $\lesssim 10$AU. However, the interplay between photoevaporation
and layered accretion can be subtle even in single-star disks
\citep{morishima12}, and detailed investigation of this issue is
beyond the scope of this paper.

The increased surface density in the dead zone (compared to a fully
turbulent model) may lead to damping of the perturbations to the
planetesimals from the binary torque. For example, for a fixed disk
aspect ratio, $H/r$, the drag force increases linearly with the
surface density \citep[e.g.][]{marzari00}.  Because the drag force is
also inversely proportional to the particle size, comparing the effect
of particle sizes is equivalent to comparing varying surface
density. The larger the particle size the larger the damping and thus
we expect the increased surface density in the dead zone to increase
the damping. The effect of the large surface density in the dead zone
on the planetesimal perturbations should be investigated further in
future work.

\section{Conclusions}

With circumbinary disk models we find a dead zone typically extends
from a radius close to the inner disk edge up to a radius of around
$10-20\,\rm au$ for a close binary. The dead zone provides a quiescent
region where solids can settle to the mid-plane, ideal for planet
formation. A peak in the surface density occurs in the dead zone, far
from the inner disk edge, and dust particles of all sizes drift towards
this radius. Thus, the binary torque, that makes the inner regions of
the disk hostile to planetesimal accretion, may not affect the growth
of planetesimals in a disk with a dead zone. The currently observed
circumbinary planets likely formed in such a region before their
inward migration to their current location. The snow line typically
occurs close to but slightly outside of the peak in the surface
density. Thus, we expect both massive gas giants and terrestrial
planets to form in such disks. Because the binary provides a torque on
the disk preventing accretion, the dead zone is not accreted on to the
binary and lasts the lifetime of the disk. Thus, planet formation may
be even more likely in circumbinary disks around close binaries than
in circumstellar disks. Dead zones around wide binaries are less
likely and thus we suggest planet formation may be more difficult
there.

\section*{Acknowledgements}

We thank an anonymous referee for useful comments.  RGM's support was
provided in part under contract with the California Institute of
Technology (Caltech) funded by NASA through the Sagan Fellowship
Program. PJA acknowledges support from NASA under grant HST-AR-12814
awarded by the Space Telescope Science Institute, which is operated by
the Association of Universities for Research in Astronomy, Inc., for
NASA, under contact NAS 5-26555, and from NASA's Origins of Solar
Systems program under grant NNX13AI58G. RDA acknowledges support from
the Science \& Technology Facilities Council (STFC) through an
Advanced Fellowship (ST/G00711X/1) and Consolidated Grant
ST/K001000/1.

\label{lastpage}
\end{document}